\documentclass[conference]{IEEEtran}
\IEEEoverridecommandlockouts
\usepackage{cite}
\usepackage{amsmath,amssymb,amsfonts}
\usepackage{algorithmic}
\usepackage{algorithm}
\usepackage{bm}
\usepackage{color}
\usepackage{lineno}
\usepackage{multirow}
\usepackage{graphicx}
\usepackage{url}
\usepackage[
colorlinks=true,
urlcolor=blue,
citecolor=blue,
linkcolor=blue,
]{hyperref}

\usepackage{my}
\RequirePackage[normalem]{ulem} 
\RequirePackage{color}\definecolor{RED}{rgb}{1,0,0}\definecolor{BLUE}{rgb}{0,0,1} 
\providecommand{\DIFaddbegin}{} 
\providecommand{\DIFaddend}{} 
\providecommand{\DIFdelbegin}{} 
\providecommand{\DIFdelend}{} 
\providecommand{\DIFaddbeginFL}{} 
\providecommand{\DIFaddendFL}{} 
\providecommand{\DIFdelbeginFL}{} 
\providecommand{\DIFdelendFL}{} 
\newcommand{\DIFscaledelfig}{0.5}
\RequirePackage{settobox} 
\RequirePackage{letltxmacro} 
\newsavebox{\DIFdelgraphicsbox} 
\newlength{\DIFdelgraphicswidth} 
\newlength{\DIFdelgraphicsheight} 
\LetLtxMacro{\DIFOincludegraphics}{\includegraphics} 
\newcommand{\DIFaddincludegraphics}[2][]{{\color{blue}\fbox{\DIFOincludegraphics[#1]{#2}}}} 
\newcommand{\DIFdelincludegraphics}[2][]{
\sbox{\DIFdelgraphicsbox}{\DIFOincludegraphics[#1]{#2}}
\settoboxwidth{\DIFdelgraphicswidth}{\DIFdelgraphicsbox} 
\settoboxtotalheight{\DIFdelgraphicsheight}{\DIFdelgraphicsbox} 
\scalebox{\DIFscaledelfig}{
\parbox[b]{\DIFdelgraphicswidth}{\usebox{\DIFdelgraphicsbox}\\[-\baselineskip] \rule{\DIFdelgraphicswidth}{0em}}\llap{\resizebox{\DIFdelgraphicswidth}{\DIFdelgraphicsheight}{
\setlength{\unitlength}{\DIFdelgraphicswidth}
\begin{picture}(1,1)
\thicklines\linethickness{2pt} 
{\color[rgb]{1,0,0}\put(0,0){\framebox(1,1){}}}
{\color[rgb]{1,0,0}\put(0,0){\line( 1,1){1}}}
{\color[rgb]{1,0,0}\put(0,1){\line(1,-1){1}}}
\end{picture}
}\hspace*{3pt}}} 
} 
\LetLtxMacro{\DIFOaddbegin}{\DIFaddbegin} 
\LetLtxMacro{\DIFOaddend}{\DIFaddend} 
\LetLtxMacro{\DIFOdelbegin}{\DIFdelbegin} 
\LetLtxMacro{\DIFOdelend}{\DIFdelend} 
\DeclareRobustCommand{\DIFaddbegin}{\DIFOaddbegin \let\includegraphics\DIFaddincludegraphics} 
\DeclareRobustCommand{\DIFaddend}{\DIFOaddend \let\includegraphics\DIFOincludegraphics} 
\DeclareRobustCommand{\DIFdelbegin}{\DIFOdelbegin \let\includegraphics\DIFdelincludegraphics} 
\DeclareRobustCommand{\DIFdelend}{\DIFOaddend \let\includegraphics\DIFOincludegraphics} 
\LetLtxMacro{\DIFOaddbeginFL}{\DIFaddbeginFL} 
\LetLtxMacro{\DIFOaddendFL}{\DIFaddendFL} 
\LetLtxMacro{\DIFOdelbeginFL}{\DIFdelbeginFL} 
\LetLtxMacro{\DIFOdelendFL}{\DIFdelendFL} 
\DeclareRobustCommand{\DIFaddbeginFL}{\DIFOaddbeginFL \let\includegraphics\DIFaddincludegraphics} 
\DeclareRobustCommand{\DIFaddendFL}{\DIFOaddendFL \let\includegraphics\DIFOincludegraphics} 
\DeclareRobustCommand{\DIFdelbeginFL}{\DIFOdelbeginFL \let\includegraphics\DIFdelincludegraphics} 
\DeclareRobustCommand{\DIFdelendFL}{\DIFOaddendFL \let\includegraphics\DIFOincludegraphics} 
\RequirePackage{listings} 
\RequirePackage{color} 
\lstdefinelanguage{DIFcode}{ 
  moredelim=[il][\color{red}\sout]{\%DIF\ <\ }, 
  moredelim=[il][\color{blue}\uwave]{\%DIF\ >\ } 
} 
\lstdefinestyle{DIFverbatimstyle}{ 
	language=DIFcode, 
	basicstyle=\ttfamily, 
	columns=fullflexible, 
	keepspaces=true 
} 
\lstnewenvironment{DIFverbatim}{\lstset{style=DIFverbatimstyle}}{} 
\lstnewenvironment{DIFverbatim*}{\lstset{style=DIFverbatimstyle,showspaces=true}}{} 

\begin{document}

\title{
GPU Optimization of Lattice Boltzmann Method with Local Ensemble Transform Kalman Filter
    \thanks{    
    This work was supported in part by JSPS KAKENHI [Grant Numbers JP21K17755, JP22H03599],
    Joint Usage/Research Center for Interdisciplinary Large-scale Information Infrastructures (JHPCN) [Project IDs jh210049, jh220030, jh220031].
    Computations were performed on the Wisteria/BDEC-01 Aquarius subsystem at The University of Tokyo.

    \copyright 2022 IEEE. Personal use of this material is permitted. Permission from IEEE must be obtained for all other uses, in any current or future media, including reprinting/republishing this material for advertising or promotional purposes, creating new collective works, for resale or redistribution to servers or lists, or reuse of any copyrighted component of this work in other works.
    }
}


\author{
\IEEEauthorblockN{Yuta Hasegawa}
\IEEEauthorblockA{\textit{Center for Computational Science \& e-Systems} \\
\textit{Japan Atomic Energy Agency}\\
Kashiwa-shi, Chiba, Japan \\
ORCID: 0000-0002-4072-6311}

\and

\IEEEauthorblockN{Toshiyuki Imamura}
\IEEEauthorblockA{\textit{Center for Computational Science} \\
\textit{RIKEN} \\
Kobe, Hyogo, Japan \\
ORCID: 0000-0003-1601-9710}

\linebreakand

\IEEEauthorblockN{Takuya Ina}
\IEEEauthorblockA{\textit{Center for Computational Science \& e-Systems} \\
\textit{Japan Atomic Energy Agency}\\
Kashiwa-shi, Chiba, Japan \\
ORCID: 0000-0002-3989-5011}

\and

\IEEEauthorblockN{Naoyuki Onodera}
\IEEEauthorblockA{\textit{Center for Computational Science \& e-Systems} \\
\textit{Japan Atomic Energy Agency}\\
Kashiwa-shi, Chiba, Japan \\
ORCID: 0000-0001-7392-2899}

\linebreakand

\IEEEauthorblockN{Yuuichi Asahi}
\IEEEauthorblockA{\textit{Center for Computational Science \& e-Systems} \\
\textit{Japan Atomic Energy Agency}\\
Kashiwa-shi, Chiba, Japan \\
ORCID: 0000-0002-9997-1274}

\and

\IEEEauthorblockN{Yasuhiro Idomura}
\IEEEauthorblockA{\textit{Center for Computational Science \& e-Systems} \\
\textit{Japan Atomic Energy Agency}\\
Kashiwa-shi, Chiba, Japan \\
ORCID: 0000-0002-2829-0498}
}

\maketitle

\begin{abstract}
The ensemble data assimilation of computational fluid dynamics simulations based on the lattice Boltzmann method (LBM) and the local ensemble transform Kalman filter (LETKF) is implemented and optimized on a GPU supercomputer based on NVIDIA A100 GPUs. To connect the LBM and LETKF parts, data transpose communication is optimized by overlapping computation, file I/O, and communication based on data dependency in each LETKF kernel. In two dimensional forced isotropic turbulence simulations with the ensemble size of $M=64$ and the number of grid points of $N_x=128^2$, the optimized implementation achieved $\times3.80$ speedup from the naive implementation, in which the LETKF part is not parallelized. The main computing kernel of the local problem is the eigenvalue decomposition (EVD) of $M\times M$ real symmetric dense matrices, which is computed by a newly developed batched EVD in \verb|EigenG|. The batched EVD in \verb|EigenG| outperforms that in \verb|cuSOLVER|, and $\times65.3$ speedup was achieved.
\end{abstract}


\thispagestyle{plain}
\pagestyle{plain}


\section{Introduction\label{sec:introduction}}

Chaotic nature of nonlinear dynamical systems makes the prediction of transient phenomena difficult without any external information from the real world.
The ensemble data assimilation (DA) such as the ensemble Kalman filters (EnKFs)~\cite{Evensen2003a,Tippett2003} is one of the promising means to introduce the observation data from the real world into the numerical simulation.
The ensemble DA has been emerged in the field of weather prediction research, and nowadays it is also applied to other fields, e.g., computational fluid dynamics~(CFD) simulations of turbulent flows~\cite{Colburn2011,Labahn2020,DeMarinis2021} and phase field simulations of the dendrite growth~\cite{Yamanaka2019,Yamanaka2020,Miyoshi2021a}.

With increasing computing power, the fidelity of simulation is dramatically improved, and the main focus of computational science is shifting to the accuracy and credibility of the simulation conditions such as the initial condition, the boundary condition, and hyper parameters in the simulation. To resolve the latter issue, high performance computing (HPC) not only of the simulation but also of the DA is becoming important. From the viewpoint of HPC, the local ensemble transform Kalman filter (LETKF)~\cite{Hunt2007a} is suitable for massively parallel implementation, and is widely utilized in recent DA problems.
The LETKF computes the $M\times M$ covariance matrix, which is transformed from model space to ensemble space, on each grid point, while in the original EnKF, the size of the covariance matrix $N\times N$ in model space is huge. 
Here, the sizes of ensemble space $M$ and model space $N$ are typically of the order of $M=O(10^2)\sim O(10^4)$ and $N\geq O(10^8)$, respectively.
The LETKF drastically reduces the size of the covariance matrix, which enables the update of each grid point in an embarrassingly parallel manner on each processor.
Thanks to this feature, the LETKF was successfully applied to extreme scale DA problems with $N\geq O(10^8)$ and $M\geq O(10^2)$ on state-of-the-art supercomputers~\cite{Miyoshi2014,Miyoshi2016a,Yashiro2020}.

Although the large scale ensemble DA with the LETKF has been realized on CPU-based supercomputers,
its implementation on GPU-based supercomputers has not been demonstrated yet.
Therefore, in this work, we study a high performance GPU implementation of the DA problem of CFD simulations based on the lattice Boltzmann method (LBM) and the LETKF on the state-of-the-art supercomputer based on NVIDIA A100 GPUs. The followings are the main contributions of this study.
\begin{enumerate}
    \item A new batched version of the eigenvalue solver \verb|EigenG|~\cite{Imamura2014a} is developed for the LETKF, which computes $M\times M$ real symmetric dense matrices in an embarrassingly parallel manner. This solver outperforms \verb|cuSOLVER|~\cite{cusolver} at $M=32$, 48, and 64, where $\times 1.71$, $\times 118$, and $\times 65.3$ speedups are achieved.
    \item Communication processes of ensemble data are optimized by taking account of data dependency in the LETKF, and the LBM-LETKF model for two dimensional (2D) turbulence simulations with the ensemble size of $M=64$ and the number of grid points of $N_x=128^2$ on 64 GPUs achieved $\times 3.80$ speedup from a naive implementation. 
\end{enumerate}

The remainder of this paper is organized as following.
\sref{sec:methodology} introduces the numerical schemes of the LBM and the LETKF.
\sref{sec:eigeng} presents the batched version of \verb|EigenG|.
\sref{sec:parallelization} explains the parallelization schemes for the LETKF.
\sref{sec:result} carries out the performance evaluation of 2D turbulence simulations with the LETKF on the GPU-based supercomputer.
Finally, \sref{sec:conclusion} concludes this study.
\section{Methodology\label{sec:methodology}}

\subsection{Lattice Boltzmann method (LBM)\label{ssec:lbm}}

CFD simulations are performed using the LBM, which is briefly summarized below.
More details of the numerical models used in this study can be found in our previous work~\cite{my:lbm2d-letkf}.
The LBM is one of the fully explicit schemes for incompressible flow problems, which were conventionally calculated by the Navier-Stokes equations including the pressure Poisson equation for an implicit time integration.
The origin of the governing equation is the Boltzmann equation, which is based on the kinetic theory in phase space (position and velocity).
In the LBM, its discretization in configuration subspace is based on uniform Cartesian grids,
and that in velocity subspace is represented by a finite number of directions.
Thanks to the fully explicit scheme, the LBM is scalable and has been applied to extreme scale problems with over $10^{10}$ degrees-of-freedoms~\cite{Onodera2013a, Randles2015}.

The lattice Boltzmann equation is expressed as
\begin{align}
    f_{ij|x+ci\Delta t, y+cj\Delta t|t+\Delta t} = f_{ij|xy|t} + \Omega_{ij|xy|t},
\end{align}
where $f_{ij|xy|t}$ is the variable called the distribution function.
$(i, j)$ are indices of velocity components, which are defined as $(i,j)\in\{-1,0,1\}\otimes\{-1,0,1\}$, i.e., the D2Q9 LBM is employed in this study.
$(x,y)$ and $\Delta x$ are the position and the grid spacing in Cartesian grids, respectively.
$t$ and $\Delta t$ are time and the time step width, respectively.
$\Omega_{ij|xy|t}: \mathbb{R}^{9}\to\mathbb{R}^{9}$ is the collision operator, which computes the collisional interaction within the nine distribution functions on each grid point.
As the collision operator, the lattice-BGK collision model~\cite{Qian1992} and the Ladd's second-order forcing term~\cite{Ladd1994} are employed.
Finally, the fluid density $\rho$ and the velocity $\bm u$ are derived by taking the moments of the distribution functions:
\begin{align}
    \rho_{xy|t} &= \sum_{i,j} f_{ij|xy|t}, \\
    \bm u_{xy|t} &= \cfrac{1}{\rho_{xy|t}} \sum_{i,j} \bm c_{ij} f_{ij|xy|t},
\end{align}
where $\bm c_{ij}$ is the discrete velocity, defined as 
\begin{align}
    \bm c_{ij} &= (ci, cj)^\top, \\
    c &= \Delta x/\Delta t.
\end{align}

\subsection{Local ensemble transform Kalman filter (LETKF)\label{ssec:letkf}}

As the ensemble DA model, we employ the LETKF~\cite{Hunt2007a}.
We firstly define the variables as follows:
\begin{itemize}
    \item $\bm x$: 
    the state vector in model space.
    \item $\bm P = (\bm x - \bm x^\mathrm{t})(\bm x - \bm x^\mathrm{t})^\top$:
    the error covariance in model space,
    where $\bm x^\mathrm{t}$ is the state vector of the ground truth.
    \item $\bm y$:
    a vector of observed values.
    \item $\bm H$:
    the observation matrix which gives $\bm y = \bm H \bm x$.
    \item $\bm R= (\bm y^\mathrm{o} - \bm H \bm x^\mathrm{t})(\bm y^\mathrm{o} - \bm H \bm x^\mathrm{t})^\top$:
    the error covariance in observation space, where the superscript `o' denotes the observation (i.e. the external information).
\end{itemize}
In the current LBM model, $\bm x$ corresponds to $f_{ji|xy|t}$ and $\bm y$ is given by $\rho_{xy|t}$ and $\bm u_{xy|t}$ at observation points.

The Kalman filter (KF) gives the linear estimation:
\begin{align}
    \label{eq:kfx}
    \bm x^\mathrm{a} &= \bm x^\mathrm{f} + \bm K \left( \bm y^\mathrm{o} - \bm H \bm x^\mathrm{f} \right),\\
    \label{eq:kfp}
    \bm P^\mathrm{a} &= (\bm I - \bm K\bm H) \bm P^\mathrm{f},\\
    \label{eq:kfk}
    \bm K &= \bm P^\mathrm{f} \bm H^\top\left( \bm H\bm P^\mathrm{f}\bm H^\top + \bm R \right)^{-1},
\end{align}
where the superscripts `f' and `a' respectively denote the variables of forecast (prior) and analysis (posterior), 
$\bm I$ is the identity matrix,
and $\bm K$ is a matrix called as the Kalman gain.
The above vectors and matrices are updated via the following DA cycles:
\begin{enumerate}
    \item Set the initial value $\bm x^\mathrm{a}|_{t=0}$ and $\bm P^\mathrm{a}|_{t=0}$.
    \item Compute time integration to obtain $\bm x^\mathrm{f}|_{t+s\Delta t}$ and $\bm P^\mathrm{f}|_{t+s\Delta t}$ from $\bm x^\mathrm{a}|_t$ and $\bm P^\mathrm{a}|_t$.
    Here, $s$ is the number of time integration sub-cycles per DA cycle.
    \item Compute DA to estimate $\bm x^\mathrm{a}|_{t+s\Delta t}$ from $\bm x^\mathrm{f}|_{t+s\Delta t}$ and $\bm y^\mathrm{o}|_{t+s\Delta t}$.
    \item Repeat 2) and 3).
\end{enumerate}

In the LETKF or any other type of EnKFs, the simulation is performed in parallel (namely, ensemble simulations), so that $\bm P$ is represented by the ensemble statistics.
The ensemble mean vector, the perturbation vector and the perturbation matrix are respectively given by
\begin{align}
    \label{eq:xmean}
    \bar{\bm x} &= \cfrac{1}{M}\sum_{m=0}^{M-1} \bm x_m, \\
    \delta {\bm x}_m &= \bm x_m-\bar{\bm x}_m, \\
    \label{eq:xspread}
    \delta\bm X &= \left[
        \delta \bm x_0, \delta \bm x_1, \ldots, \delta \bm x_{M-1}
    \right],
\end{align}
where $\bm x_m$ is the state vector of $m$-th ensemble member. Hereafter, we use the same convention as Eq. (\ref{eq:xspread}) in writing ensemble vectors in a matrix form.
The error covariance $\bm P$ is then approximately estimated as
\begin{align}
    \label{eq:enkf_p}
    \bm P &= \cfrac{1}{M-1}\sum_{m=0}^{M-1} 
        \left( \bm x_m - \bar{\bm x} \right)
        \left( \bm x_m - \bar{\bm x} \right)^\top
        \nonumber \\
    &= \cfrac{1}{M-1}\delta \bm X \delta \bm X^\top.
\end{align}
Here, the state vector of the ground truth is approximated by the ensemble mean ($\bm x^\mathrm{t}=\bar{\bm x}$), as the ground truth state is not observable.
These expressions of EnKFs in model space are further reduced by transforming them into ensemble space (for detail, cf. e.g.~\cite{Li2010book}).
The analysis vector is transformed as
\begin{align}
    \label{eq:letkf_xa}
    \bm x^\mathrm{a}_m &= \bar{\bm x}^\mathrm{f} 
        + \delta \bm X^\mathrm{f}\left( \bar{\bm w}^\mathrm{a} + \delta \bm w^\mathrm{a}_m \right),
\end{align}
where $\bar{\bm w}^\mathrm{a}$ and $\delta \bm w^\mathrm{a}_m$ are the mean and perturbation of the transformed vector, respectively.
They are solved by the following equations:
\begin{align}
    \label{eq:letkf_wa}
    \bar{\bm w}^\mathrm{a} &= \tilde{\bm P}^\mathrm{a} \delta \bm Y^{\mathrm{f}\top} \bm R^{-1} \left( \bm y^\mathrm{o} - \bar{\bm y}^\mathrm{f} \right), \\
    \label{eq:letkf_dwa}
    \delta \bm W^\mathrm{a} &= \sqrt{M-1} (\tilde{\bm P}^\mathrm{a})^{1/2},
\end{align}
where
\begin{align}
    \tilde{\bm P}^\mathrm{a} &= \bm Q^{-1}, \\
    \label{eq:letkf_q}
    \bm Q &= (M-1)\bm I + \delta \bm Y^{\mathrm{f}\top} \bm R^{-1} \delta \bm Y^{\mathrm{f}},\\
    \label{eq:enkf_ybar}
    \bar{\bm y}^\mathrm{f} &= \cfrac{1}{M}\sum_{m=0}^{M-1} \bm y^{\mathrm{f}}_m,\\
    \label{eq:enkf_deltay}
    \delta\bm y^\mathrm{f}_m &= \bm y^\mathrm{f}_m - \bar{\bm y}^\mathrm{f}.
\end{align}
Here, $\tilde{\bm P}^\mathrm{a}$ and $(\tilde{\bm P}^\mathrm{a})^{1/2}$ in Eqs. \eqref{eq:letkf_wa} and \eqref{eq:letkf_dwa} are computed by the eigenvalue decomposition (EVD) of $\bm Q$.
The size of $\bm Q$ is $M\times M$, and thus, the problem size of the EVD varies depending on the ensemble size, while it is not affected by the numbers of grid points, variables or observation points.

By applying the $\bm R$-localization model~\cite{Hunt2007a}, the above equations are decomposed into a local problem on each grid point. 
In the $\bm R$-localization model, the inverse of the observation error covariance $\bm R^{-1}$ is replaced by $\bm R_\mathrm{loc}^{-1}=\bm G \circ\bm R^{-1}$, where $\bm G$ is a diagonal matrix given by a localization function. 
In this work, the localization function is given by the Gaspari-Cohn function~\cite{Gaspari1999}, in which the cutoff distance is chosen as $d=2(p-1)\Delta x$ based on the spacing of observation points $p\Delta x$.
In the original global problem, the sizes of the state vector $\bm x$, the observation vector $\bm y$, and the transformed vector $\bm w$ are $N=N_xN_v$, $LV$, and $M$, respectively, where $N_x$ and $N_v=9$ are respectively the number of grid points in configuration and velocity subspaces, $V$ is the number of observed variables, and $L$ is the number of observation points. 
On the other hand, the local problem on each grid point is computed using $N_v$ elements of $\bm x$, $lV$ elements of $\bm y$, and $\bm w$, where $l$ is the number of local observation points within the cutoff distance from the grid point. 
It is noted that in the current LBM model, the local problem is defined at each grid point in configuration subspace, because the observation data, $\rho_{xy|t}, \bm u_{xy|t}$ and the localization function do not depend on the velocity $\bm c_{ij}$ and nine velocity grids share the same observation data.

The computational procedures of the LETKF are summarized as follows.
\begin{enumerate}
    \item Compute the forecast observation data $\mathsf{Y}^\mathrm{f}[M,LV]$ from the forecast ensemble data $\mathsf{X}^\mathrm{f}[M,N]$ and the observation matrix $\mathsf{H}[LV,N]$.
    \item Compute the local observation data on each grid point $\mathsf{Y}_\mathrm{loc}^\mathrm{f}[M, N_x, l_\mathrm{max}V]$ using the $\bm R$-localization model. Here, the size of the local observation data is defined by the maximum number of $l$, which changes depending on the alignment between the grid point and the surrounding observation points.
    \item Compute the transformed vector $\mathsf{W}^\mathrm{a}[M,M]$ by solving the local problem, and update the analysis ensemble data $\mathsf{X}^\mathrm{a}[M, N]$.
\end{enumerate}
Here, $\mathsf{X}^\mathrm{f}[M,N]$, $\mathsf{Y}^\mathrm{f}[M,LV]$, $\mathsf{H}[LV,N]$, $\mathsf{W}^\mathrm{a}[M,M]$, and $\mathsf{X}^\mathrm{a}[M,N]$ are multi-dimensional arrays corresponding to $\bm X^\mathrm{f}$, $\bm Y^\mathrm{f}$, $\bm H$, $\bm W^\mathrm{a}$, and $\bm X^\mathrm{a}$, respectively. 
The indices in the brackets denote the size of array in the row-major format. 
$\mathsf{Y}_\mathrm{loc}^\mathrm{f}[M, N_x, l_\mathrm{max}V]$ is an array of the local observation data.

By using the above notation, the overall procedures of the LETKF are described in \algoref{algo:base}.
In this algorithm, the local problem on each grid point can be computed in parallel, where each local problem consists of small matrix computations. This kind of bulk matrix computation is suitable for batched matrix solvers.
In this work, we utilize the vendor-provided library \verb|cuBLAS|~\cite{cublas} for the batched GEMM operations, and the in-house library \verb|EigenG| (shown in \sref{sec:eigeng}) for solving batched EVDs.

\begin{algorithm}[t]
    \caption{Basic implementation of the LETKF.\label{algo:base}}
    \begin{algorithmic}[1]
        \REQUIRE $\mathsf{X}^\mathrm{f}[M, N]$ from ensemble simulation
\REQUIRE $\mathsf{Y}^\mathrm{o}[LV]$ from ground truth observation data
\ENSURE $\mathsf{X}^\mathrm{a}[M, N]$
\STATE Compute $\mathsf{Y}^\mathrm{f}[M, LV]$ $\Leftarrow$ $\mathsf{X}^\mathrm{f}[M, N],\mathsf{H}[LV, N]$
\STATE Extract local observation data: \\
    $\mathsf{Y}^\mathrm{o}_\mathrm{loc}[N_x, l_\mathrm{max}V]$ $\Leftarrow$ $\mathsf{Y}^\mathrm{o}[LV]$; \\
    $\mathsf{Y}^\mathrm{f}_\mathrm{loc}[M, N_x, l_\mathrm{max}V]$ $\Leftarrow$ $\mathsf{Y}^\mathrm{f}[M, LV]$
\FOR{$n_x\in[0,N_x)$}
\FOR{$n_v\in[0,N_v)$}
    \STATE $n=N_vn_x+n_v$
    \STATE Compute local problem:    \\
    $\mathsf{W}^\mathrm{a}[M, M]\Leftarrow \mathsf{Y}^\mathrm{f}_\mathrm{loc}[M, n_x, l_\mathrm{max}V], \mathsf{Y}^\mathrm{o}_\mathrm{loc}[n_x, l_\mathrm{max}V]$\\
    $\mathsf{X}^\mathrm{a}[M, n]\Leftarrow \mathsf{X}^\mathrm{f}[M, n], \mathsf{W}^\mathrm{a}[M, M]$
\ENDFOR
\ENDFOR
    \end{algorithmic}
\end{algorithm}

\section{Batched eigenvalue solver\label{sec:eigeng}}

Among the existing eigenvalue solvers on GPUs, the batched EVD is supported only in \verb|cuSOLVER|~\cite{cusolver}.
The batched EVD \verb|cusolverDnSsyevjBatched| is an iterative solver based on the Jacobi method, and supports efficient batched computation up to the matrix size of $M=32$. To overcome this limitation, we extend our eigenvalue solver \verb|EigenG|~\cite{Imamura2014a} to a batched version \verb|EigenGBatched|. The original \verb|EigenG| was designed for relatively large matrices, and the standard eigenvalue computation algorithm consisting of the Sorrensen-Dongarra's block Householder tridiagonalization, the eigenvalue computation via Cuppen's divide and conquer algorithm, and the eigenvector computation via the block Householder back-transformation was used. However, in the batched version, we target small matrices, and employ the Sorrensen-Dongarra's block Householder tridiagonalization, the implicit-shift QL approach adopted in \verb|EISPACK|~\cite{Garbow1974} and \verb|LAPACK|~\cite{lapack99}, and the Householder back-transform that uses Joffrain's block reflector~\cite{Joffrain2006}.

The GPU implementation uses a method, in which a single warp is employed to process a single batch. In addition, since the workload in the warp for problems with $M<32$ is unbalanced in each CUDA core, cooperative groups (tiled partitioning) are adopted according to the size of the target matrices.
Specifically, when $M\ge 32$, one-batch per warp is adopted, and when $1<M<32$, $32/T$ batches are assigned to a single warp,
where the tile size is defined as $T=2^{\mathrm{int}(\mathrm{log}_2(M-1)+1)}$.
In the implementation, three primitive APIs for inter-thread collective operations are provided; 
1) intra-group synchronization (syncwarp functionality), 2) intra-group reduction (implemented by butterfly sum with warp-shuffle instructions), and 3) intra-group broadcast (implemented with warp-shuffle instructions). 
When the vector length exceeds the warp length, multiple elements are allocated to one thread in a simple round-robin fashion. 
In the cooperative group, the three collectives can be provided in an almost similar format, and thus, there is no difference between the former and latter implementations except for the loop for allocating multiple vector elements to each CUDA core. 
In addition, each implementation is optimized by efficiently using the memory hierarchy consisting of registers, shared memory, and device memory and by avoiding warp divergence via the use of a ternary operator for a conditional branch.




The performance comparison between \verb|cusolverDnSsyevjBatched| and \verb|EigenGBatched| is conducted for the batched matrix data from the LBM-LETKF model with the batch size of $N_x=128^2$ and the ensemble size of $M=16$, 32, 48, and 64, where each batch computes EVD of $M\times M$ real symmetric dense matrix in FP32. The computational costs on a single A100 GPU are summarized in Table \ref{tab1}. At $M=16$, \verb|cusolverDnSsyevjBatched| is $\times1.86$ faster than \verb|EigenGBatched|. However, at $M=32$, 48, and 64, \verb|EigenGBatched| outperforms \verb|cusolverDnSsyevjBatched|, where the performance gains are respectively $\times1.71$, $\times118$, and $\times65.3$.
In both solvers, the computational cost increases with the matrix size. However, they behave differently beyond the threshold matrix size of $M=32$. 
The computational cost of \verb|cusolverDnSsyevjBatched| shows an explosive growth of $\times225$ between $M=32$ and $M=48$, while the computational cost is almost saturated between $M=48$ and $M=64$. 
On the other hand, that of \verb|EigenGBatched| shows a modest growth of $\times3.25$ between $M=32$ and $M=48$. The theoretical complexity of the current EVD is $O(M^3)$, which may be affected by the following two effects. One is an additional overhead for the round-robin allocation of multiple elements for $M>32$, and the other is an initialization overhead, which becomes relatively small at larger $M$. Since the cost increase of $\times3.25$ is less than the ideal scaling of $\times3.375$, the latter cost reduction dominates over the former overhead. 

{
The performance difference between \verb|EigenGBatched| and \verb|cusolverDnSsyevjBatched| depends largely on the numerical algorithms employed. The peak ratios of the memory bandwidth observed by the visual profiler are 63\% and 38\% for \verb|EigenGBatched| and \verb|cusolverDnSsyevjBatched|, respectively, for the matrix size of $M=32$, suggesting that their performance difference is attributed to the workload tuning. In addition, when the matrix size is larger than $M=32$, the \verb|cusolverDnSsyevjBatched| calls different non-batched kernels by $N_x$ times, and thus, every batch is processed sequentially, leading to significant performance degradation. 
}

The eigenvalue computation, such as the implicit-shift QL, contains an iterative method internally, and its behavior depends on the distribution of eigenvalues. This leads to significantly different behavior from group to group or thread to thread, and the warp divergent situation that is of concern in CUDA is thought to be occurring. In order to achieve higher efficiency, it is necessary to explore algorithms that can avoid warp divergence and approaches that directly compute the square root of matrices. These issues will be addressed in the future.

\begin{table}[tb]
\caption{Performance comparison between batched EVDs in cuSOLVER and EigenG. The batched matrix data in FP32 generated from the LBM-LETKF model with the batch size of $N_x=128^2$ and the ensemble size of $M=16$, 32, 48, and 64 are computed on a single A100 GPU. The elapsed time is shown in msec.}
\begin{tabular}[t]{l|ccccc}
\hline
matrix size & \verb|cusolverDnSsyevjBatched| & \verb|EigenGBatched| \\
\hline
$16\times 16$ & 0.7 & 1.3 \\
$32\times 32$ & 4.1 & 2.4 \\
$48\times 48$ & 922.8 & 7.8 \\
$64\times 64$ & 927.4 & 14.2 \\
\hline
\end{tabular}
\label{tab1}
\end{table}
\section{Parallelization of LETKF\label{sec:parallelization}}

In the current LBM-LETKF model, the LBM model for each ensemble member is computed using a single GPU, and the number of GPUs and thus MPI processes is chosen to be the same as the ensemble size $M$. This is a typical parallelization strategy in our ensemble simulations, while the resource per ensemble member may change depending on the target problem. 
After computation of the LBM part, the forecast ensemble data $\mathsf{X}^\mathrm{f}[M, N]$ is stored over multiple GPUs, and $m$-th GPU keeps $\mathsf{X}^\mathrm{f}[m, N]$.
In computing the LETKF part, one need to gather or transpose the ensemble data depending on the parallelization strategy of the LETKF part. In the following, we discuss different parallelization approaches from the viewpoint of the memory size, the communication cost, and the computational cost.

\subsection{Naive implementation\label{ssec:parallelization:naive}}

We firstly show a naive implementation in \algoref{algo:letkf_prallel_naive}, in which all GPUs keep the same copy of all ensemble data via MPI\_Allgather.
This approach needs quite large memory which is $M$ times larger than the distributed data $\mathsf{X}^\mathrm{f}[m, N]$ for each ensemble member, and requires the redundant computation of the LETKF on all GPUs.
This approach may be available up to the grid size of $N=O(10^5)$ and the ensemble size of $M=O(10^2)$ due to the limited memory on each GPU.
Although such inefficiency and memory limitation exist, this approach is the simplest way to introduce the LETKF to the existing simulations. In this work, we use this implementation as the baseline for performance comparisons.

It should be noted that in the current implementation, data transpose is needed between the LBM part and the LETKF part.
In the LBM part, we use the structure-of-array (SoA) memory layout, $\mathsf{X}^\mathrm{f}[m, N_v, N_x]$, for efficient stencil computation in configuration subspace. 
On the other hand, in the LETKF part, it is converted into the array-of-structure (AoS) memory layout, $\mathsf{X}^\mathrm{f}[m, N_x, N_v]$, because on each grid point, the local problem for different $\bm c_{ij}$ is computed at once using the same local observation data.
Another data transpose is needed after MPI\_Allgather, because batched matrix computation in \verb|cuBLAS| and \verb|EigenG| assume that the $N_x$-axis (batch dimension) is outermost, $\mathsf{X}^\mathrm{f}[N_x, M, N_v]$, while MPI\_Allgather gives ensemble data with the $M$-axis (ensemble/process dimension) being outermost, $\mathsf{X}^\mathrm{f}[M, N_x, N_v]$.
However, the cost of such data transpose on a single GPU is relatively small and negligible.

\subsection{Parallel implementation\label{ssec:parallelization:distributed}}

In order to parallelize the LETKF part, $N_x$ batched tasks or the local problems are distributed over $M$ GPUs.
For such data distribution, we replace MPI\_Allgather by MPI\_Alltoall as follows.
After processing the LBM part, we convert $m$-th forecast data in the SoA layout, $\mathsf{X}^\mathrm{f}[m, N_v, N_x]$, into the AoS layout, $\mathsf{X}^\mathrm{f}[m, N_x,N_v]$.
We then execute MPI\_Alltoall($\mathsf{X}^\mathrm{f}[m, N]$) to store a part of the total forecast data, $\mathsf{X}^\mathrm{f}[M, (N/M)_m]$, on the $m$-th GPU, where $(N/M)_m$ denotes the distributed batch for $n\in[mN/M, (m+1)N/M)$ ($n_x\in[mN_x/M, (m+1)N_x/M)$ and $n_v\in[0, N_v)$), where $n_x$ and $n_v$ are the indices in the $N_x$ and $N_v$ axes, respectively, and $n=n_xN_v+n_v$.
We apply the same all-to-all communication also to the local observation data MPI\_Alltoall($\mathsf{Y}^\mathrm{f}_\mathrm{loc}[m, N_x, l_\mathrm{max}V]$).
Finally, we compute a part of the LETKF over $N_x/M$ grid points on each GPU, and execute MPI\_Alltoall($\mathsf{X}^\mathrm{a}[M, (N/M)_m]$) so that the $m$-th analysis data, $\mathsf{X}^\mathrm{a}[m, N]$, is returned on the $m$-th GPU.
These procedures are summarized in \algoref{algo:letkf_prallel_opt1}.

\begin{algorithm}[t]
    \caption{Naive parallelization of LETKF\label{algo:letkf_prallel_naive}}
    \begin{algorithmic}[1]
\REQUIRE $\mathsf{X}^\mathrm{f}[m, N]$ from simulation on $m$-th GPU
\REQUIRE $\mathsf{Y}^\mathrm{o}[LV]$ from ground truth observation data
\ENSURE $\mathsf{X}^\mathrm{a}[m, N]$
    \STATE Compute $\mathsf{Y}^\mathrm{f}[m, LV]$ $\Leftarrow$ $\mathsf{X}^\mathrm{f}[m, N],\mathsf{H}[LV, N]$
    \STATE Extract local observation data: \\
    $\mathsf{Y}^\mathrm{o}_\mathrm{loc}[N_x, l_\mathrm{max}V]$
    $\Leftarrow$
    $\mathsf{Y}^\mathrm{o}[LV]$;\\
    $\mathsf{Y}^\mathrm{f}_{loc}[m, N_x, l_\mathrm{max}V]$
    $\Leftarrow$
    $\mathsf{Y}^\mathrm{f}[m, LV]$
    \STATE $\mathsf{Y}^\mathrm{f}_\mathrm{loc}[M, N_x, l_\mathrm{max}V]$ 
    $\leftarrow$ MPI\_Allgather($\mathsf{Y}^\mathrm{f}_\mathrm{loc}[m, N_x, l_\mathrm{max}V]$)
    \STATE $\mathsf{X}^\mathrm{f}[M, N]$ $\leftarrow$ MPI\_Allgather($\mathsf{X}^\mathrm{f}[m, N]$)
\FOR{$n_x\in[0,N_x)$}
\FOR{$n_v\in[0,N_v)$}
    \STATE $n=N_vn_x+n_v$
    \STATE Compute local problem:    \\
    $\mathsf{W}^\mathrm{a}[M, M]\Leftarrow \mathsf{Y}^\mathrm{f}_\mathrm{loc}[M, n_x, l_\mathrm{max}V], \mathsf{Y}^\mathrm{o}_\mathrm{loc}[n_x, l_\mathrm{max}V]$\\
    $\mathsf{X}^\mathrm{a}[M, n]\Leftarrow \mathsf{X}^\mathrm{f}[M, n], \mathsf{W}^\mathrm{a}[M, M]$
\ENDFOR
\ENDFOR
    \end{algorithmic}
\end{algorithm}

\begin{algorithm}[t]
    \caption{Distributed parallelization of LETKF\label{algo:letkf_prallel_opt1}}
    \begin{algorithmic}[1]
\REQUIRE $\mathsf{X}^\mathrm{f}[m, N]$ from simulation on $m$-th GPU
\REQUIRE $\mathsf{Y}^\mathrm{o}[LV]$ from ground truth observation data
\ENSURE $\mathsf{X}^\mathrm{a}[m, N]$
    \STATE Compute $\mathsf{Y}^\mathrm{f}[m, LV]$ $\Leftarrow$ $\mathsf{X}^\mathrm{f}[m, N],\mathsf{H}[LV, N]$
    \STATE Extract local observation data: \\
    $\mathsf{Y}^\mathrm{o}_\mathrm{loc}[N_x, l_\mathrm{max}V]$
    $\Leftarrow$
    $\mathsf{Y}^\mathrm{o}[LV]$;\\
    $\mathsf{Y}^\mathrm{f}_\mathrm{loc}[m, N_x, l_\mathrm{max}V]$
    $\Leftarrow$
    $\mathsf{Y}^\mathrm{f}[m, LV]$    
    \STATE $\mathsf{Y}^\mathrm{f}_\mathrm{loc}[M, (N_x/M)_m, l_\mathrm{max}V]$ \\
    $\leftarrow$ MPI\_Alltoall($\mathsf{Y}^\mathrm{f}_\mathrm{loc}[m, N_x, l_\mathrm{max}V]$)
    \STATE $\mathsf{X}^\mathrm{f}[M, (N/M)_m]$ $\leftarrow$ MPI\_Alltoall($\mathsf{X}^\mathrm{f}[m, N]$)
\FOR{$n_x\in[mN_x/M,(m+1)N_x/M)$}
\FOR{$n_v\in[0,N_v)$}
    \STATE $n=N_vn_x+n_v$
    \STATE Compute local problem:    \\
    $\mathsf{W}^\mathrm{a}[M, M]\Leftarrow \mathsf{Y}^\mathrm{f}_\mathrm{loc}[M, n_x, l_\mathrm{max}V], \mathsf{Y}^\mathrm{o}_\mathrm{loc}[n_x, l_\mathrm{max}V]$\\
    $\mathsf{X}^\mathrm{a}[M, n]\Leftarrow \mathsf{X}^\mathrm{f}[M, n], \mathsf{W}^\mathrm{a}[M, M]$
\ENDFOR
\ENDFOR
    \STATE $\mathsf{X}^\mathrm{a}[m, N]$ $\leftarrow$ MPI\_Alltoall($\mathsf{X}^\mathrm{a}[M, (N/M)_m]$)
    \end{algorithmic}
\end{algorithm}

\subsection{Overlapping communication and file I/O\label{ssec:parallelization:overlap_comm}}

\begin{figure}[t]
    \centering
    \includegraphics[width=\linewidth]{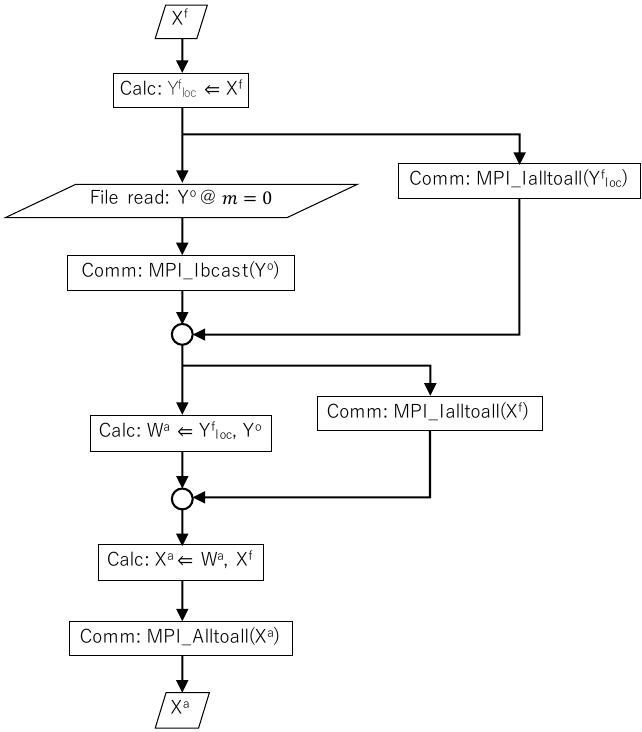}
    \caption{Data flow with overlap of calculation, communication and file I/O in LETKF.}
    \label{fig:comm_ovlp}
\end{figure}

We then consider the overlap of computation, file I/O, and communication in \algoref{algo:letkf_prallel_opt1}.
We decompose the LETKF part into the following steps:
\begin{enumerate}
    \item Compute local observation data, $\mathsf{Y}^\mathrm{f}_\mathrm{loc}$.
    \item MPI\_Alltoall($\mathsf{Y}^\mathrm{f}_\mathrm{loc}$).
    \item Read observation data, $\mathsf{Y}^\mathrm{o}$, from data file.
    \item MPI\_Alltoall($\mathsf{X}^\mathrm{f}$).
    \item Compute transformed vector, $\mathsf{W}^\mathrm{a}$, using Eqs. \eqref{eq:letkf_wa}--\eqref{eq:enkf_deltay}.
    \item Update analysis vector, $\mathsf{X}^\mathrm{a}$, using Eq. \eqref{eq:letkf_xa}.
    \item MPI\_Alltoall($\mathsf{X}^\mathrm{a}$).
\end{enumerate}
Here, 3) is the file I/O, which requires no communication data, and thus, can be overlapped with 2) or 4). 5) is a part of the LETKF, which does not depend on the forecast ensemble data from 4).
As per such data dependency, we overlap communication steps 2) and 4) with 3) and 5), respectively.
In 3), the cost of the file I/O is relatively large, and 5) includes computation of the batched EVD, which is the most costly part and occupies $\sim 50\%$ of the computational cost of the LETKF. Therefore, these communication costs are expected to be well hidden.

Moreover, we consider optimizing the file I/O.
Since the observation data is common in every ensemble member, each member reads the same data file in the original implementation.
It may lead a conflict of the file I/O between multiple GPUs, and cause the increase of the waiting time for the file I/O.
To avoid such a conflict, we optimize the file I/O using the broadcast communication.
Only the primary process ($m=0$) read the observation data and then, MPI\_Bcast($\mathsf{Y}^\mathrm{o}$) is executed.
This approach may be efficient because the bandwidth of the communication (PCI-e, NVLink and/or Infiniband) is much faster than the throughput of the file I/O.

In the communication overlap, communication steps are executed on the host memory using asynchronous routines (MPI\_Ialltoall and MPI\_Ibcast).
The remaining non-overlapped communication step 7) is implemented on the GPU memory using CUDA-aware MPI.
Detailed design of the communication overlap is summarized in \figref{fig:comm_ovlp}.

\section{Performance evaluation\label{sec:result}}

\begin{figure}[t]
    \centering
    \includegraphics[width=\linewidth]{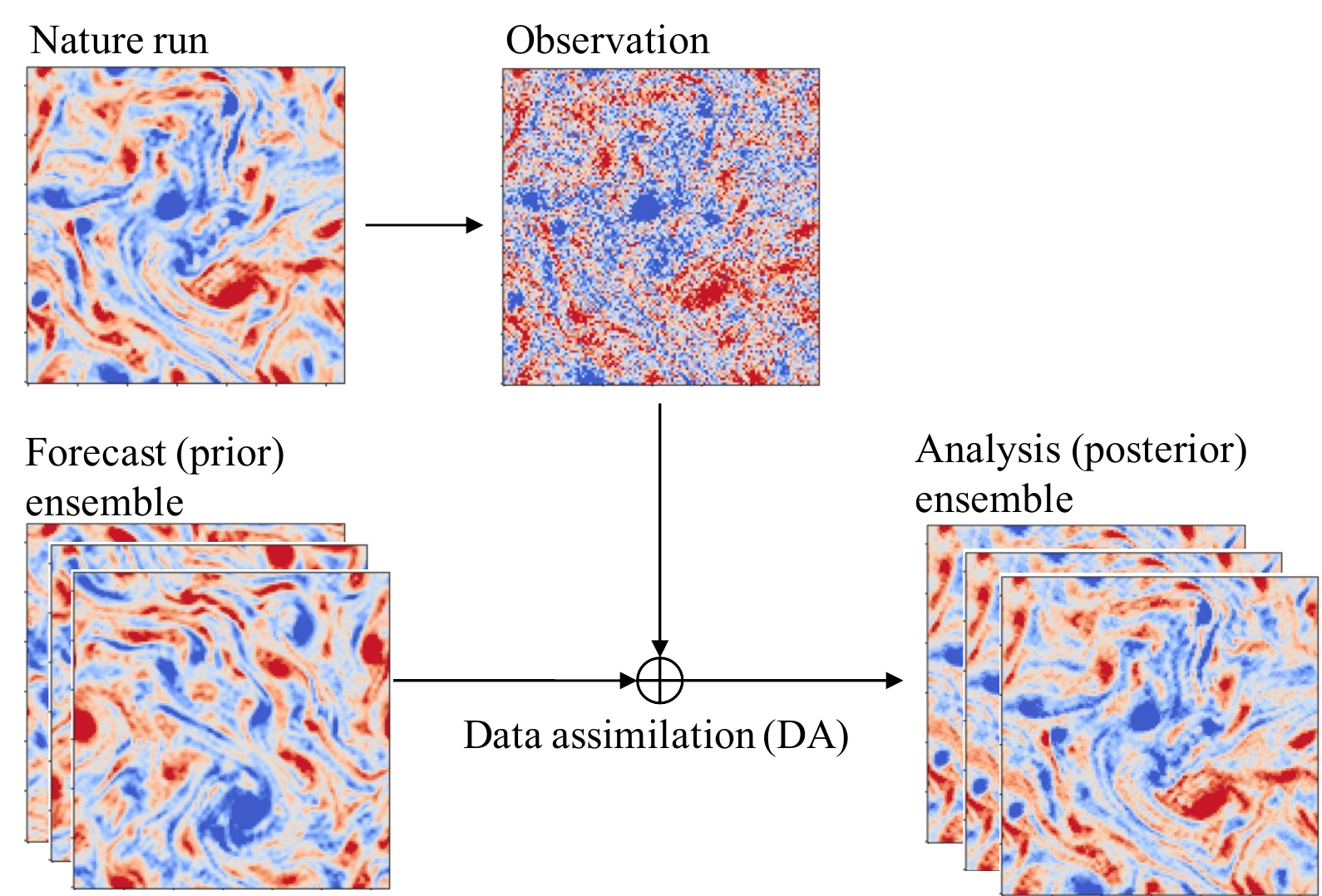}
    \caption{Schematics of the concept of the ensemble data assimilation experiment with the observing system simulation experiment (OSSE) manner.
        \label{fig:osse_concept}
        }
\end{figure}

The performance evaluation of the LBM-LETKF model is conducted for 2D forced isotropic turbulence simulations.
The state vector is given by the distribution function, $f_{ij|xy|t}$, with $N_x=128^2$ and $N_v=9$.
The observation vector consists of the macroscopic variables, the density $\rho_{xy|t}$ and the fluid velocity $\bm u_{xy|t}$, and the number of observed variables is $V=3$.
The ensemble size is chosen to be $M = 4$, 16, and 64.

The DA numerical experiments are carried out in the observing system simulation experiment (OSSE) manner as shown in \figref{fig:osse_concept}, which assumes the ground-truth being the simulation with a certain initial condition (called the nature run). 
2D forced isotropic turbulence is characterized by an inverse energy cascade, which generates large scale vortices, and we reproduce the dynamics of such vortices in the nature run via the LETKF with noisy observation data at a limited number of observation points. 
The observation data of the nature run is stored on the storage in advance. 
The forecast ensemble data are produced by ensemble simulations, where the initial conditions are chosen randomly.
By combining the observation data and the forecast ensemble data, the LETKF outputs the analysis ensemble data, which modify the state vector towards the nature run.
The observation matrix is defined to add $\sim 20\%$ of the Gaussian observation noise, and its resolution is chosen to be spatially dense (in \sref{ssec:result:dense}) or sparse (in \sref{ssec:result:sparse}). 
The number of time integration sub-cycles per DA cycle or the DA interval is set to $s=100$, which is determined based on the speed of the error growth (i.e. the Lyapunov exponent)~\cite{Labahn2020}.

The DA numerical experiments are performed on the Aquarius system at the University of Tokyo~\cite{Aquarius}.
The Aquarius system consists of NVIDIA A100 GPU 40GB SXM (19.5 TFLOPs in FP32, 1555 GB/s; 8 GPUs per node), Intel Xeon Platinum 8360Y CPU (2.4 GHz, 36 cores; 2 CPUs per node), and the inter-node communication via the Infiniband HDR (200 Gbps $\times$4 per node). The intra-node GPU-GPU communication is supported by NVLink (600 GB/s per GPU). The storage system consists of DDN SFA7990XE (504GB/s).
As the compilers and the MPI library, we utilize \verb|GCC| (8.3.1), \verb|CUDA| (11.2), and \verb|OpenMPI| (4.1.1, CUDA-aware MPI enabled).

\subsection{Dense observation case\label{ssec:result:dense}}

\begin{figure}[t]
    \centering
    \includegraphics[width=\linewidth]{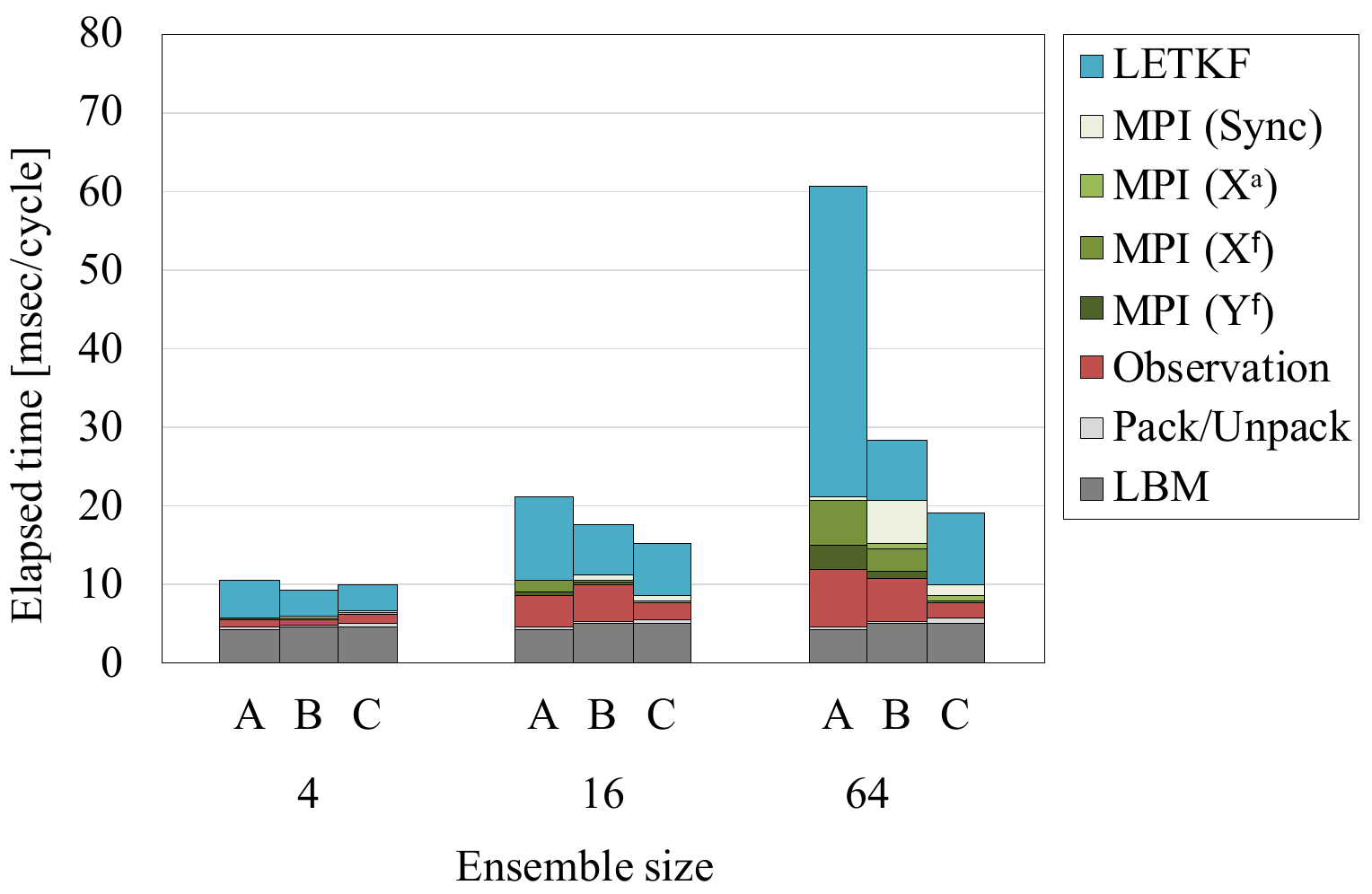}
    \caption{Elapsed time on spatially-dense observation test (in Sec.~\ref{ssec:result:dense}).
        `A', `B' and `C' denote the subsection numbers which refer the parallelization schemes described in \sref{sec:parallelization}.
        Legend denotes the breakdown of the elapsed time.
        \texttt{LBM}: Calculating time integral of the LBM.
        \texttt{Pack/Unpack}: Data packing/unpacking between LBM and LETKF, data transposing after collective communications.
        \texttt{Observation}: Loading observation data.
        \texttt{MPI(X$^\texttt{f}$), MPI(Y$^\texttt{f}$), MPI(X$\texttt{a}$)}: Executing collective communication for $\bm X^\mathrm{f}$, $\bm Y^\mathrm{f}$ and $\bm X^\mathrm{a}$.
        \texttt{MPI(Sync)}: Synchronization overhead.
        \texttt{LETKF}: Solving LETKF, including batched computation of EVD, and BLAS operations such as GEMM and GEMV.
        }
    \label{fig:result_dense}
\end{figure}

In the dense observation case, fully local observation with $l_\mathrm{max}=1$ is performed for every grid point, $L=N_x=128^2$. This condition gives the most accurate DA solution, where the root mean square error (RMSE) of $\bm u_{xy|t}$ was a few percent~\cite{my:lbm2d-letkf}.
\figref{fig:result_dense} shows the performance comparisons of the naive implementation (\sref{ssec:parallelization:naive}, case A), the parallel implementation (\sref{ssec:parallelization:distributed}, case B), and the parallel implementation with optimized communication (\sref{ssec:parallelization:overlap_comm}, case C). 
At $M=4$, all three cases show similar performances, because the matrix size of $M=4$ is too small to fill out GPU threads, and the cost of intra-node communication via NVLink is negligibly small. 
However, with increasing $M$, cases B and C show significant performance gains from case A.
At $M=64$, the most part of the elapsed time is occupied by \texttt{MPI($\cdot$)} (steps 2), 4), and 7)) and \verb|LETKF| (steps 5) and 6)), and \verb|LBM| and \verb|Observation| (steps 1) and 3)) give minor contributions.
The computational cost of \verb|LETKF| is reduced from 39.5 msec/cycle (case A) to 8.85 msec/cycle (case C), leading to $\times5.30$ speedup.
Here, the batch size per GPU is reduced to $1/M=1/64$. However, the above speedup is smaller than the ratio of the batch size per GPU.
This result suggests that the batch size per GPU ($N_x/M=16384/64=256$) is too small for the batched EVD.
The cost of \texttt{MPI($\cdot$)} is also reduced from 8.84 msec/cycle (case A) to 4.34 msec/cycle (case B), leading to $\times2.04$ speedup.
Moreover, in case C, the communication costs of $\mathsf{X}^\mathrm{f}$ and $\mathsf{Y}^\mathrm{f}_\mathrm{loc}$ are respectively hidden behind \verb|LETKF| and \verb|Observation|, and the remaining communication cost is reduced to 0.94 msec/cycle ($\times4.64$ speedup from case B).
By using broadcast communication, the cost of \verb|Observation| is also reduced from 5.51 msec/cycle (case B) to 1.96 msec/cycle (case C).
Overall, the total performance including both the LBM and LETKF parts shows $\times$3.20 speedup between case A and case C.

\subsection{Sparse observation case\label{ssec:result:sparse}}

\begin{figure}[t]
    \centering
    \includegraphics[width=\linewidth]{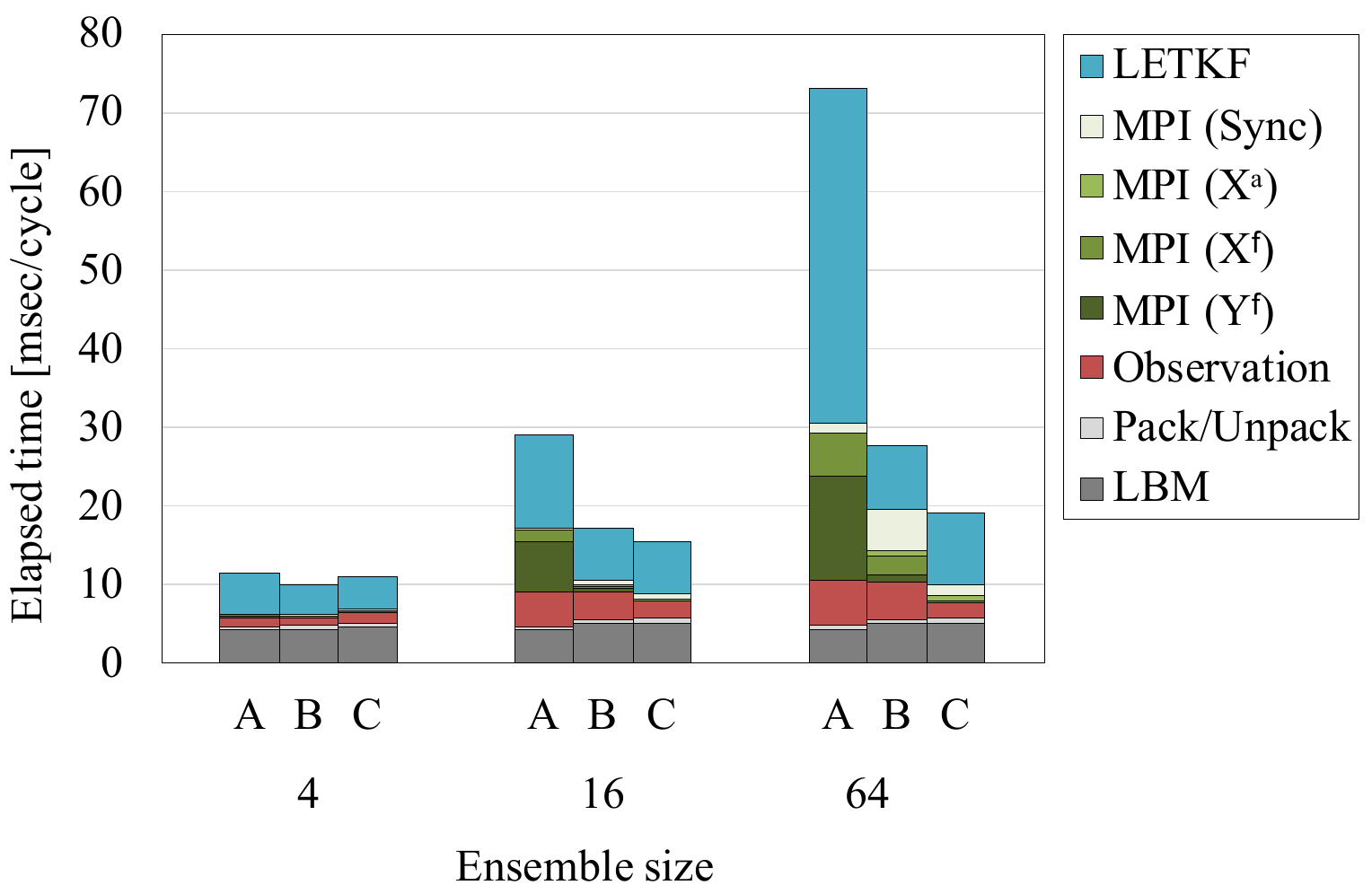}
    \caption{Elapsed time on spatially-sparse observation test (in Sec.~\ref{ssec:result:sparse}).
        `A', `B' and `C' denote the subsection numbers which refer the parallelization schemes described in \sref{sec:parallelization}.
        The legend marks same meaning as \figref{fig:result_dense}.
        \label{fig:result_sparse}
        }
\end{figure}

In the sparse observation case, the resolution of the observation is chosen as $L=32^2$, which is 6.25\% of $N_x$. 
The observation points are located on every four grid points in configuration subspace, $p=4$. 
In this observation condition, the RMSE of $\bm u_{xy|t}$ was suppressed below 10\%~\cite{my:lbm2d-letkf}. 
It is noted that the sparse observation case often becomes numerically unstable because of poor resolution and statistics of the observation data. 
Therefore, there exists a trade-off between the spacing of observation points $p$ and the ensemble size $M$, and the LETKF with less observation points requires more ensembles.
As the cutoff distance is given by $d=2(p-1)\Delta x=6\Delta x$, the maximum number of local observation points becomes $l_{\rm max}=4^2$, leading to $\times 16$ increase of the local observation data, $\mathsf{Y}^\mathrm{f}_\mathrm{loc}$, from the dense observation case.
This data size is $\times 5.3$ larger than that of the state vectors, $\mathsf{X}^\mathrm{f}$ or $\mathsf{X}^\mathrm{a}$.

\figref{fig:result_sparse} shows the performance comparisons in the sparse observation case.
Compared with the dense observation case, the communication cost of \texttt{MPI($\mathsf{Y}^\mathrm{f}_\mathrm{loc}$)} is significantly increased in cases A and B.
However, in case C, this cost is still well hidden behind \texttt{Observation}. 
Except for this step, the remaining steps are almost unchanged from \sref{ssec:result:dense}, and similar performance gains are obtained in cases B and C. Consequently, the total performance at $M=64$ achieves $\times3.80$ speedup between the naive implementation (case A, 73.0 msec/cycle) and the optimal one (case C, 19.2 msec/cycle). It is noted that thanks to the communication overlap, case C gives similar costs between the dense and sparse observation cases. This feature is of critical importance, because in the real DA problems, the observation condition is normally sparse, and less observation points are preferable from the viewpoint of the experimental cost.




\section{Conclusion\label{sec:conclusion}}

In this study, we optimized the GPU implementation of the LBM-LETKF model. 
In the LBM part, each ensemble member is computed independently on each GPU. On the other hand, in the LETKF part, the analysis ensemble data is computed using the forecast ensemble data generated from the LBM part. As the LETKF requires all ensemble data, one needs to gather or transpose the ensemble data before the LETKF part. In the naive implementation, all GPUs gather all ensemble data via MPI\_Allgather. Although this approach is quite simple, large memory space and redundant computation of the LETKF are needed. In this work, we developed a parallel implementation, in which the LETKF part is also parallelized by transposing the ensemble data via MPI\_Alltoall. In addition, by analyzing the data dependency in the LETKF, the parallel implementation is further optimized by overlapping computation, file I/O, and communication. In 2D forced isotropic turbulence simulations with the ensemble size of $M=64$ and the number of grid points of $N_x=128^2$, the optimized implementation achieved $\times 3.80$ speedup from the naive implementation. 

In the LETKF part, the LETKF is decomposed into a local problem on each grid point. The local problem consists of small matrix computation with the matrix size of $M\times M$, and can be computed using batched matrix operations in an embarrassingly parallel manner. The batched operations of the LETKF are implemented using \verb|cuBLAS| and \verb|EigenG|. Here, a new batched EVD is developed based on Sorrensen-Dongarra's block Householder tridiagonalization, the implicit-shift QL approach, and the Householder back-transform. The batched EVD in \verb|EigenG| outperforms that in \verb|cuSOLVER|, and at $M=64$, $\times 65.3$ speedup was achieved.

\bibliographystyle{IEEEtran}

\end{document}